\documentclass[]{elsart}
\usepackage{natbib}
\usepackage{epsfig}

\def\aap{\ {Astron. \& Astrophys.}\ }

\def\aj{\ {Astron.\, J.}\ }

\def\apj{\ {Astrophys.\, J.}\ }

\def\apss{\ {Ap\&SS}\ }
\def\araa{\ {ARA\&A}\ }

\def\mnras{\ {Mon.\, Not.\, R.\, Astron.\, Soc.}\ }
\def\nat{\ {Nature}\ }

\def\pasp{\ {PASP}\ }
\def\pasj{\ {Publ. Astr. Soc. Japan}\ }

\newcommand\kira{{\tt kira}}
\newcommand\tpr{{\tau_{\rm pr}}} 
\newcommand\tcr{{\tau_{\rm cr}}} 

\newcommand\jerk{{\mbox{${\bf k}$}}}

\def\apgt{\ {\raise-.5ex\hbox{$\buildrel>\over\sim$}}\ }
\def\aplt{\ {\raise-.5ex\hbox{$\buildrel<\over\sim$}}\ }

\begin{document}
\begin{frontmatter}

%[HPC $N$-body simulations on GPU and GRAPE]
\title{A parallel gravitational $N$-body kernel}

\author[SCS,API]{Simon Portegies Zwart}
\author[Drexel]{Stephen McMillan}
\author[SCS,API]{Derek Groen}
\author[RIT]{Alessia Gualandris}
\author[Dwingeloo]{Michael Sipior}
\author[SARA]{Willem Vermin}

\address[SCS]{Section Computational Science, University of Amsterdam, 
         Amsterdam, The Netherlands}
\address[API]{Astronomical Institute "Anton Pannekoek" , University of 
         Amsterdam, Amsterdam, The Netherlands}
\address[Drexel]{Department of Physics, Drexel University, Philadelphia, 
         PA 19104, USA}
\address[Dwingeloo]{ASTRON, Oude Hoogeveensedijk 4, 7991 PD Dwingeloo}
\address[RIT]{Rochester Institute of Technology, 85 Lomb Memorial Drive, Rochester, NY 14623, USA}
\address[SARA]{SARA Computing and Networking Services,
	 Kruislaan 415, 1098 SJ, Amsterdam, the Netherlands}

\begin{abstract}
We describe source code level parallelization for the {\tt kira}
direct gravitational $N$-body integrator, the workhorse of the {\tt
starlab} production environment for simulating dense stellar systems.
The parallelization strategy, called ``j-parallelization'', involves
the partition of the computational domain by distributing all
particles in the system among the available processors.  Partial
forces on the particles to be advanced are calculated in parallel by
their parent processors, and are then summed in a final global
operation. Once total forces are obtained, the computing elements
proceed to the computation of their particle trajectories.  We report
the results of timing measurements on four different parallel
computers, and compare them with theoretical predictions.  The
computers employ either a high-speed interconnect, a NUMA architecture
to minimize the communication overhead or are distributed in a grid.
The code scales well in the domain tested, which ranges from 1024 -
65536 stars on 1 - 128 processors, providing satisfactory speedup.
Running the production environment on a grid becomes inefficient for
more than 60 processors distributed across three sites.

\end{abstract}

\end{frontmatter}

%----------------------------------------------------------------------
\section{Introduction}

Numerical studies of gravitational $N$-body problem have made enormous
advances during the last few decades due to algorithmic design
\citep{1963ZA.....57...47V,1964ApNr....9..313A,1968BAN....19..479V,1975ARA&A..13....1A,1999PASP..111.1333A},
developments in hardware
\citep{1986LNP...267...86A,1996IAUS..174..141T,1998sssp.book.....M,2001ASPC..228...87M,2003PASJ...55.1163M},
and refinements to approximate methods
\citep{1971Ap&SS..13..284H,1979ApJ...234.1036C,1984ApJ...283..801M,1985ApJ...299....1B,1986Natur.324..446B}.
A number of interesting books have been written on the subject
\citep{Newton:1687,1987gady.book.....B,1987degc.book.....S,Hockney1988,2003gmbp.book.....H,Aarseth2003}.

The main reasons for the popularity of
the direct evaluation techniques are its very high accuracy and
lack of prior assumptions.  Approximate methods, by contrast,
often deal with idealized circumstances, which, regardless of their
advances in performance, may be hard to justify. 
Regardless of all the advances over the last 40 years the
evaluation of the mutual forces between all stars remains the
bottleneck in the calculation. Without the use of
hybrid force evaluation techniques, as pioneered by
\cite{1993ApJ...414..200M,1997AGAb...13...53H}, the $\mathcal{O}(N^2)$
performance complexity will remain a problem for the forseeable
future. Nevertheless, it is to be expected that the direct
integration method for gravitational $N$-body simulations will remain
popular for the next decade.

The main technique adopted to increase the performance (and accuracy)
of direct $N$-body methods in stellar dynamics applications is
individual time stepping, providing orders of magnitude increase in
computational speed.  In the early 1990s this technique enabled
astronomers to study star clusters with up to $\sim 1000$ stars
\cite{1990ApJ...362..522M,1991ApJ...372..111M,1994ApJ...427..793M}.
The main development in the 1990s was the development of
special-purpose hardware to provide the required teraflop/s raw
performance. In particular GRAPE-4 \citep{1996IAUS..174..141T}, which
enables simulation of several tens of thousands of stars, and GRAPE-6
which raised the bar above $10^5$ stars for the first time
\citep{2003PASJ...55.1163M}. For the forseeable future, special
purpose hardware may still take the lead in advancing the study of
self gravitating systems, particularly if that hardware is
multi-purpose \citep{2005PASJ...57..799H} and therefore able to serve
a larger group of users and investors.  The use of graphical processor
units (GPU), is an interesing parallel development with a promising
performance/price ratio
\citep{2007NewA...12..641P,2007arXiv0707.0438B,2007astro.ph..3182H}. Another
likely developmental pathway is the more general use of large parallel
computers.  The latter, however, requires efficient parallel kernels
for force evaluation.

In the last decade the number of processors per computer has increased
dramatically. Even today's personal computers are equipped with
multiple cores, and large supercomputers often carry many thousands of
processors, for example the IBM BlueGene/L which has 131072
processors.  Efficient interprocess communication for such large
clusters is far from trivial, and requires significant development
efforts.  The communication overhead and development challenges
increase even more once large calculations are performed on the grid,
as shown in the pioneering study of
\cite{2007PARCO..33..3G,2007arXiv0709.4552G}.

The parallelization of direct $N$-body kernels, however, has not kept
pace with the increased speed of multiprocessor computers.  Whereas
the long-range of the gravitational interaction introduces the need to
perform full $N^2$ force evaluations, it also requres all-to-all
memory access between processors in the computer, giving a large
communication overhead. This overhead increases linearly with the
number of processors \cite{2007NewA...12..357H}, although algorithms
with reduced communication characteristics have been proposed
\cite{2002NewA....7..373M}. In fact, many algorithmic advances, such
as individual time steps
\citep{1991ApJ...369..200M,1993ApJ...414..200M} and neighbor schemes
\cite{1973ApJ...179..885A}, appear to hamper the parallelization of
direct $N$-body codes.  The first efficiently parallelized production
$N$-body code was NBODY6++ \citep{2007MNRAS.374..703K}, for which the
systolic-ring algorithm was adopted
\citep{Dorband,2007PARCO..33..3G,2007NewA...12..357H}.

In this research note we report on our endevour to parallelize the
{\tt kira} direct gravitational $N$-body integrator, the workhorse of
the {\tt starlab} environment. The source code of the package,
including the parallelized version, here reported, is available online
at {\tt http://www.manybody.org/}. We now briefly report on the
implementation of the gravitational $N$-body methodology
(\S\,\ref{Sect:Integrator}) and the adopted parallelization topology
(\S\,\ref{Sect:parallelization}). In \S\,\ref{Sect:performance} we
report the performance of the package when ported to several parallel
systems.

%----------------------------------------------------------------------
\section{The gravitational $N$-body simulations}
\label{Sect:Integrator}

\subsection{Calculating the force and integrating the particles}

The gravitational evolution of a system consisting of $N$ stars with
masses $m_j$ and positions ${\bf r}_j$ is computed by the direct
summation of the Newtonian force between all pairs of stars.  The
force ${\bf F}_i$ on particle $i$ is obtained by summation
over the other $N-1$ particles:
\begin{equation}
  {\bf F}_i \equiv m_i {\bf a}_i =     m_i {\mathcal G} \sum^{N}
                   _{j=1, j \ne i} 
                   m_j 
                  {{\bf r}_i-{\bf r}_j \over |{\bf r}_i-{\bf r}_j|^3},
\label{Eq:Force}\end{equation}
where ${\mathcal G}$ is the gravitational constant.

A cluster consisting of $N$ stars evolves dynamically due to the
mutual gravity of the individual stars. For the force calculation on
each star, a total of ${1 \over 2} N(N-1)$ partial forces have to be
computed.  The resultant $\mathcal{O}(N^2)$ operation is the
bottleneck for the gravitational $N$-body problem.  Several
approximate techniques have been designed to speedup the force
evaluation, but these cannot compete with brute force in terms of
precision.

An alternative to improve the performance is by parallelizing the
force evaluation for use on a Beowulf or linux cluster (with or
without dedicated hardware)\citep{2007NewA...12..357H}; the use of
graphical processing units \citep{2007NewA...12..641P}; a large
parallel supercomputer \citep{Dorband,2007NewA...12..357H}; or for
grid operations \citep{2007PARCO..33..3G,2007arXiv0709.4552G}.  For
distributed hardware it is crucial to implement an algorithm that
limits communication as much as possible, otherwise the bottleneck
shifts from the force evaluation to interprocessor communication.

The parallelization scheme described in this paper is implemented in
the {\tt kira} $N$-body integrator, which is a part of the {\tt
starlab} package \citep{2001MNRAS.321..199P}.  In \kira\, the particle
motion is calculated using a fourth-order, individual time step
Hermite predictor-corrector scheme (Makino and Aarseth
1992).\nocite{1992PASJ...44..141M} This scheme works as follows:
During a time step the positions (${\bf x}$) and velocities (${\bf v}
\equiv \dot{{\bf x}}$) are first predicted to fourth order using the
acceleration (${\bf a} \equiv \ddot{{\bf x}}$) and the ``jerk''
($\jerk \equiv \dot{{\bf a}}$) which are known from the end of the
previous step.

The predicted position (${\bf x}_p$) and velocity (${\bf v}_p$) are
calculated for all particles
\begin{eqnarray}
        {\bf x}_p &=& {\bf x} + {\bf v} dt
	                      + {1 \over 2} {\bf a} dt^2
	                      + {1 \over 6} \jerk dt^3, \\
        {\bf v}_p &=& {\bf v} + {\bf a} dt
	                      + {1 \over 2} \jerk dt^2. 
\end{eqnarray}

The acceleration (${\bf a}_p$) and jerk ($\jerk_p$) are then
recalculated at the predicted time from $x_p$ and $v_p$ using direct
summation. Finally, a correction is based on the estimated
higher-order derivatives:
\begin{eqnarray}
        \ddot{\bf a} &=& -6 \Delta {\bf a}/dt^2 - (4 \jerk + 2\jerk_p)/dt,\\
        \ddot{\jerk} &=& 12 \Delta {\bf a}/dt^3 + 6(\jerk + \jerk_p)/dt^2. 
\end{eqnarray}
Here $\Delta {\bf a} = {\bf a} - {\bf a}_p$. 
Which then leads to the new position and velocity at time $t+dt$. 
\begin{eqnarray}
        {\bf x} &=& {\bf x}_p + {\ddot{\bf a} \over 24}dt^4  
                              + {\ddot{\jerk} \over 120} dt^5, \\
        {\bf v} &=& {\bf v}_p + {\ddot{\bf a} \over 6 } dt^3 
	                      + {\ddot{\jerk} \over 24 } dt^4.
\end{eqnarray}
The value of ${\bf a}$ and $\jerk$ are computed by direct summation.

The new timestep is calculated using a new predicted second derivative
of the acceleration $\ddot{\bf a}_p = \ddot{\bf a} + \ddot{\jerk} dt$
for each particle $i$ individually with \citep{1985mts..conf..377A}
\begin{equation}
	dt =	\left( \nu {
	               |{\bf a}_p| |\ddot{\bf a}_p| + \jerk^2 
                       \over
                       |\jerk| |\ddot{\jerk}|+\ddot{\bf a}_p^2 }
                \right)^{1/2}.
\label{Eq:timestep}
\end{equation}
Here we use for accuracy parameter $\nu = 0.01$.

A single integration step in the integrator thus proceeds as follows:
\begin{itemize}
\item[$\bullet$] Determine which stars are to be updated. Each star
      has associated with it an individual time ($t_i$) at which it
      was last advanced, and an individual time step ($dt_i$). The
      list of stars to be integrated consists of those with the
      smallest $t_i+dt_i$.  Time steps are constrained to be powers of
      2, allowing ``blocks'' of many stars to be advanced
      \citep{1991ApJ...369..200M,1993ApJ...414..200M}.
\item[$\bullet$] Before the step is taken, check for system
     reinitialization, diagnostic output, escape removal, termination
     of the run, storing data, etc.
\item[$\bullet$] Perform low-order prediction of all particles to the
      new time $t_i+dt_i$. This operation may be performed on the
      GRAPE, if present.
\item[$\bullet$] Recompute the acceleration and jerk on all stars in
      the current block (using the GRAPE or GPU, if available), and
      correct their positions and velocities.
\item[$\bullet$] Check for and initiate unperturbed motion.
\item[$\bullet$] Check for close encounters and stellar collisions.
\item[$\bullet$] Check for reorganisations in the data structure.
\item[$\bullet$] Apply stellar and/or binary evolution, and correct
      the dynamics as necessary. A more detailed discussion of
      Starlab's stellar and binary evolution packages may be found in
      \cite{1996A&A...309..179P}.
\end{itemize}

\subsection{The data structure}\label{Sect:datastructures}

An N-body system in Starlab is represented as a linked-list structure,
in the form of a mainly ``flat'' tree, individual stars being the
``leaves.'' The tree is flat in the sense that single stars
(i.e. stars that are not members of any multiple system) are
represented by top-level nodes, having the root node as
parent. Binary, triple, and more complex multiple systems are
represented as binary trees below their top-level center of mass
nodes. The tree structure determines both how node dynamics is
implemented and how the long-range gravitational force is
computed. The motion of every node relative to its parent is followed
using the Hermite predictor-corrector scheme described above. The use
of relative coordinates at every level ensures that high numerical
precision is maintained at all times, even during very close
encounters without the need for KS-regularization \citep{KS1965}.

\begin{figure}
\psfig{figure=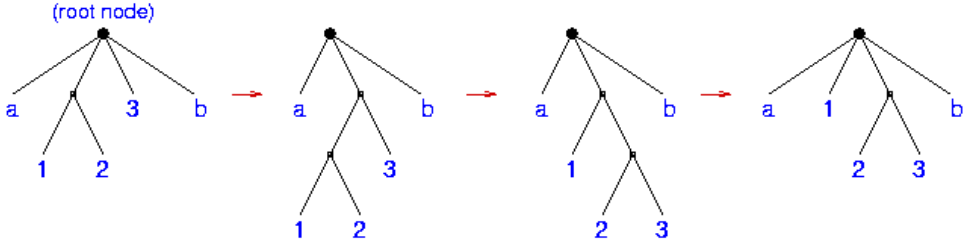,width=\columnwidth}
   \caption[]{
The tree evolves dynamically according to simple heuristic rules:
particles that approach ``too close'' to one another are combined into
a center of mass and binary node; and when a node becomes ``too
large'' it is split into its binary components. These rules apply at
all levels of the tree structure, allowing arbitrarily complex systems
to be followed. The terms ``too close'' and ``too large'' are defined
by command-line variables, as described below. The tree rearrangement
corresponding to a simple three-body ``exchange'' encounter.
\label{fig:strategy} 
}
\end{figure}

How the acceleration (and jerk) on a particle or node is computed
depends on its location in the tree. Top-level nodes feel the force
due to all other top-level nodes in the system. Forces are computed
using direct summation over all other particles in the system; no tree
or neighbor-list constructs are used. (The integrator is designed
specifically to allow efficient computation of these forces using
GRAPE or GPU hardware, if available.)  Nearby binary and multiple
systems are resolved into their components, as necessary.

The internal motion of a binary component is naturally decomposed into
two parts: (1) the dominant contribution due to its companion, and (2)
the perturbative influence of the rest of the system. (Again, this
decomposition is applied recursively, at all levels in a multiple
system.) Since the perturbation drops off rapidly with distance from
the binary center of mass, in typical cases only a few near neighbors
are significant perturbers of even a moderately hard binary. These
neighbors are most efficiently handled by maintaining lists of
perturbers for each binary, recomputed at each center of mass step,
thereby greatly reducing the computational cost of the perturbation
calculation.

\begin{figure}
\psfig{figure=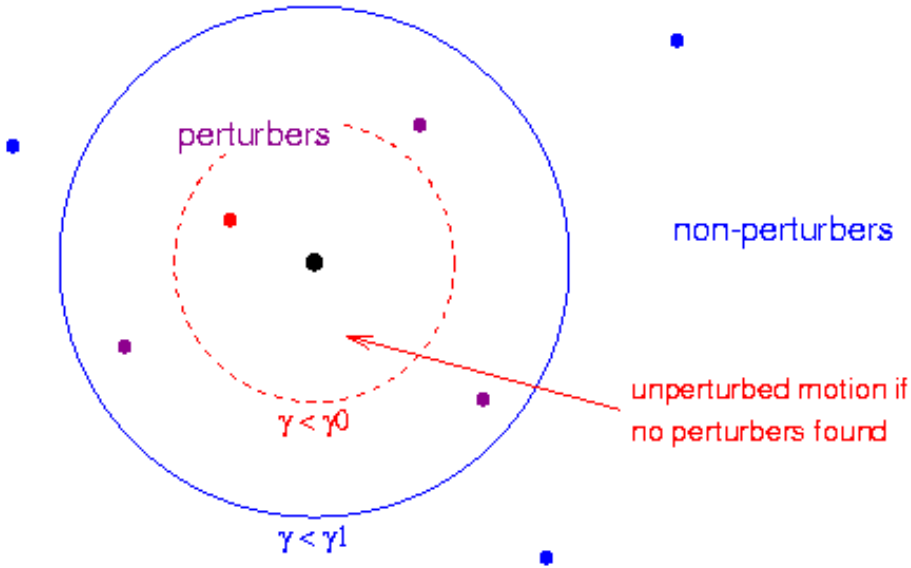,width=\columnwidth}
   \caption[]{
Illustration of the perturber treatment.
\label{fig:perturbers} 
}
\end{figure}
   
A further efficiency measure is the imposition of unperturbed motion
for binaries whose perturbation falls below some specified
value. Unperturbed binaries are followed analytically for many orbits
as strictly two-body motion; they are also treated as point masses,
from the point of view of their influence on other stars. Because
unperturbed binaries are followed in time steps that are integer
multiples of the orbit period, we relax the perturbation threshold for
unperturbed motion, relative to that for a perturbed step, as
illustrated in fig.\,\ref{fig:perturbers}. Perturbed binaries are
resolved into their components, both for purposes of determining their
center of mass motion and for determining their effect on other
stars. Unperturbed treatments of multiple systems also are used, based
on empirical studies of the stability of their internal motion.

%----------------------------------------------------------------------
\section{Parallelization strategy}\label{Sect:parallelization}

We have parallelized the above scheme (see \S\,\ref{Sect:Integrator})
by allowing each processor to compute the force between a subset of
the top-level node particles (so called $j$-parallelization). In order
to guarantee the integrity of the data across processors and to be
able to deal efficiently with neighbor lists we maintain a copy of the
entire system on each processor.

\begin{figure}
\psfig{figure=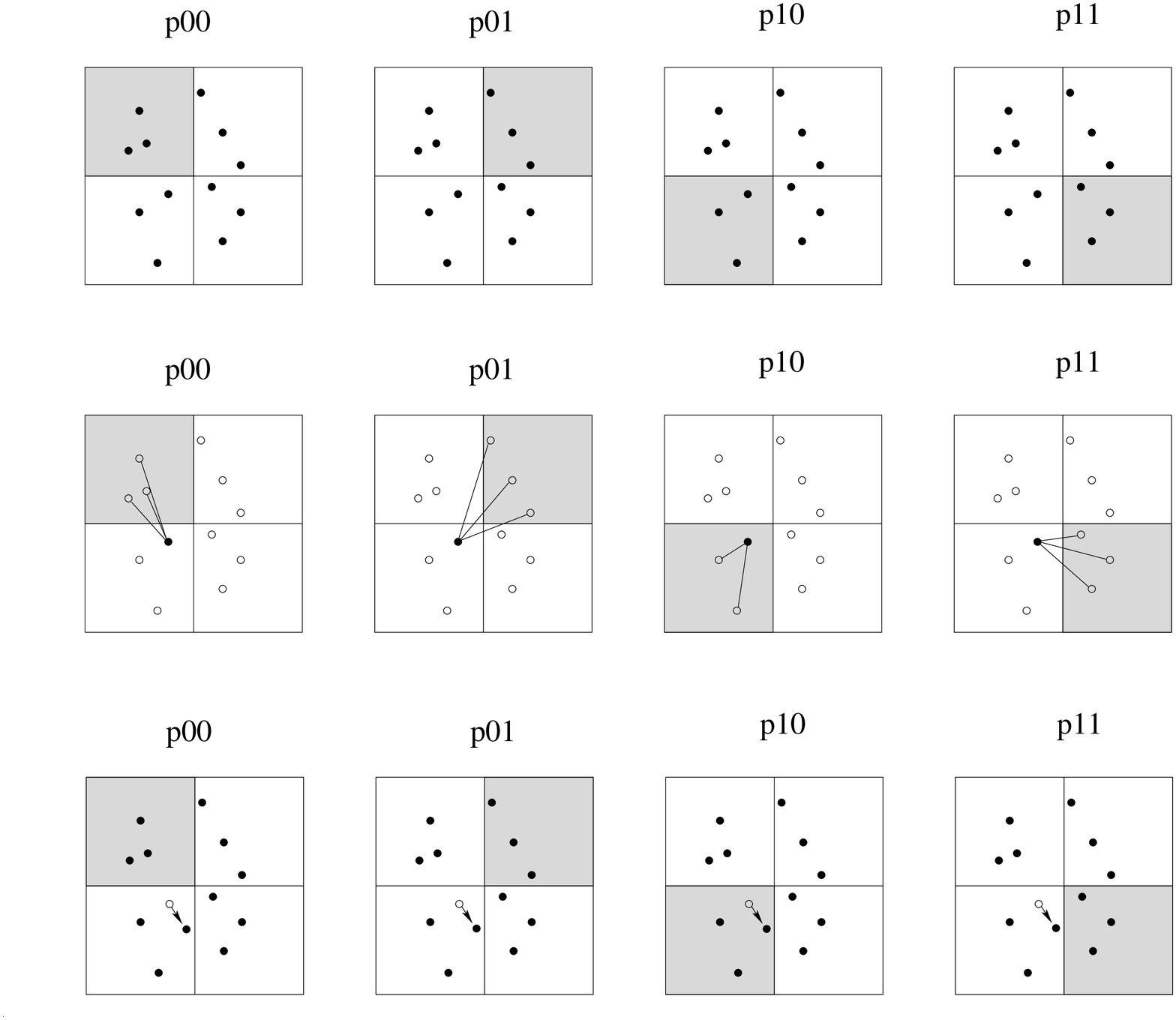,width=\columnwidth}
   \caption[]{
Strategy for the parallelization.  The top row shows the memory
management of four processors, named P00, P01, P10 and P11.  The dots
indicate the stellar system with 12 particles. Each processor has an
entiry copy of the system, but a subset $N/p$ is local (indicated with
the gray shaded area). The top row shows the memory before a step.  In
the second row one particle requires a force in order to integrate
its equations of motion. Each processor now computes the force
between this particle and its own subset.  After this step the
forces are added across the network, and each processor computes the
new position and velocity of the particle (shown in the third row).
\label{fig:strategy} 
}
\end{figure}

In fig.\,\ref{fig:strategy} we illustrate the parallalization strategy
in the case of a distributed memory computer with 4 processors, named
P00, P01, P10 and P11.  Each processor holds a complete copy of the
system in memory, but only a subset (typically $N/p$) is used by the
local processor for force calculations.  The subsets of particles are
indicated in fig.\,\ref{fig:strategy} as the grey area for each
processor.  In order to integrate the equations of motion for a
particular particle, its total force with all other particles needs to
be computed first.  Each processor then performs the force calculation
between the particle in need for an update and the set of locally
owned $N/p$ particles.  Rather than the usual $N-1$ force evaluations,
each processor then performes $(N-1)/p$ force evaluations if the
updated particle is part of its local subset of particles, and $N/p$
force evaluations if it is not local.  The partial forces are
subsequently added across the network; all processes keep thus the
same view of the system and as a consequence load balance is
guaranteed.  After the total force on a particle is calculated, each
processor calculates its new position and velocity.  The pseudo-code
for this operation is shown in the Flowchart.

\vskip0.5truein
\centerline{Flowchart}
{\small
\begin{verbatim}
  Read system
  Copy system to all processors
  Identify locally owned subset of particles for each processor
  do
    Identify particle(s) needing a force evaluation
    for all processors:
      calculate force between each particle and local particles  
    across network:
      sum all partial forces.
    calculate new position and velocity
  until the end of the simulation
\end{verbatim}\label{Tab:flowchart}
}

\subsection{Parallelizing the perturber list} \label{Sect:neighbors}

For the efficient integration of strongly perturbed single stars,
binaries and higher-order multiples, for each star the $N$-body
integrator maintains a list of stars that have the strongest effect on
its motion.  Each particle, therefore, has a linked list of
perturbers, consisting of stars which exert a force of $\apgt 10^{-7}
/\langle |{\bf F}| \rangle$ (see \S\,\ref{fig:perturbers}). Generally,
perturbers are geometrically close to the perturbed star, but this is
not necessarily the case.  Given that the force is proportional to
mass, a very massive object can still excert a considerable
perturbation even if it is far away.  We could have decided to
carefully select which stars should be local to a processor making the
following special treatment for perturbers unnecessary, but the
heuristics would be non-trivial and may depend quite sensitively on
the problem studied.

In principle it is not necessary to parallelize the perturber
treatment, but parallelizing the perturber list warrants consistency
between the parallel and the sequential implementation of the code.
Parallelization of the perturber list is hard because the order in
which the perturbers are identified must be the same in the serial
version of the code as in the parallel version, to guarantee identical
results when compared to the sequential code.  It will turn out that
the parallelization of the perturber treatment gives rise to a
performance hit for simulations with many binaries on large parallel
clusters.  In the serial version the pertuber list is filled in the
order by which perturbers are identified, until a maximum $N_{\rm
pert}$ is reached (in practice $N_{\rm pert} = 256$). We prevent
round-off for propagating through the integrator by maintaining the
same order in the perturber list on the parallel version of the
code. This is achieved by broadcasting the entire perturber list for
each particle, and then sorting them to the same order as the serial
code would have identified them.

%----------------------------------------------------------------------
\section{Performance analysis} 
\label{Sect:performance}

We have tested {\tt pkira} on four quite distinct parallel computers.
For initial conditions we selected a \cite{1911MNRAS..71..460P}
density profile with from 1024 to 65536 equal-mass particles. We
integrate this system for one eighth of an $N$-body time unit
\citep{1986LNP...267..233H}\footnote{see also {\tt
http://en.wikipedia.org/wiki/Natural\_units\#N-body\_units}.}.

\begin{figure}
\psfig{figure=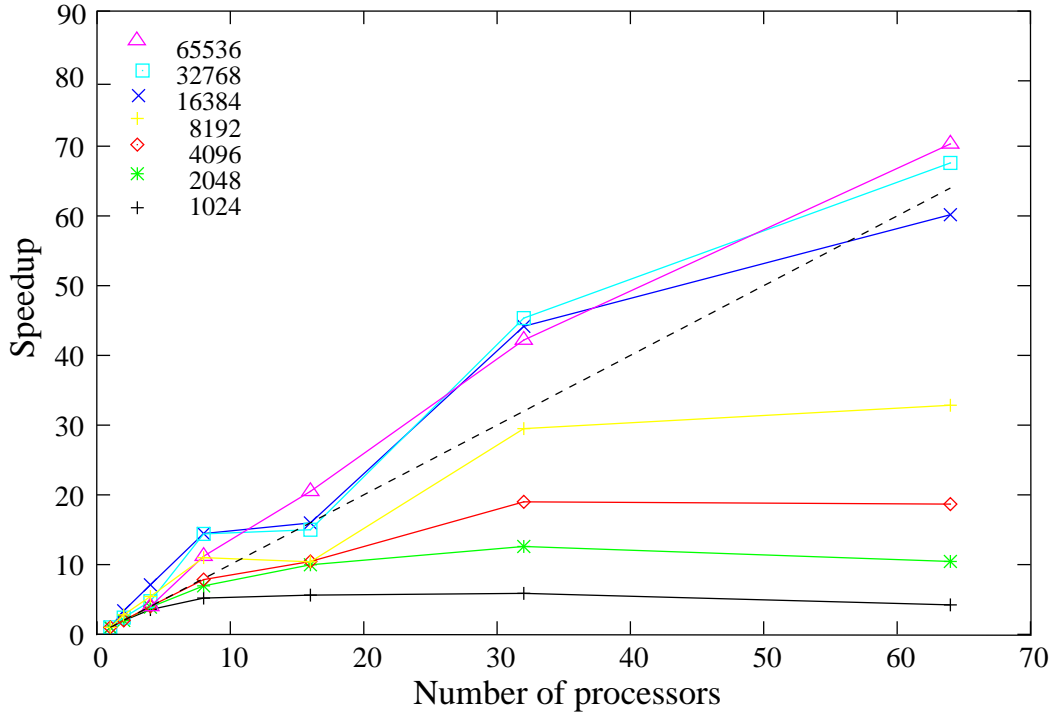,width=\columnwidth}
   \caption[]{
 Speedup for the ASTER simulated shared memory supercomputer as a
 function of the number of processors.  The various lines represent
 different particle numbers (see the legend in the top left
 corner). The diagonal dashed line give the ideal speedup. For $p
 \aplt 16$ the speedup of the code is super-ideal for $N \apgt 16384$.
\label{fig:speedup_ASTER} }
\end{figure}

\begin{figure}
\psfig{figure=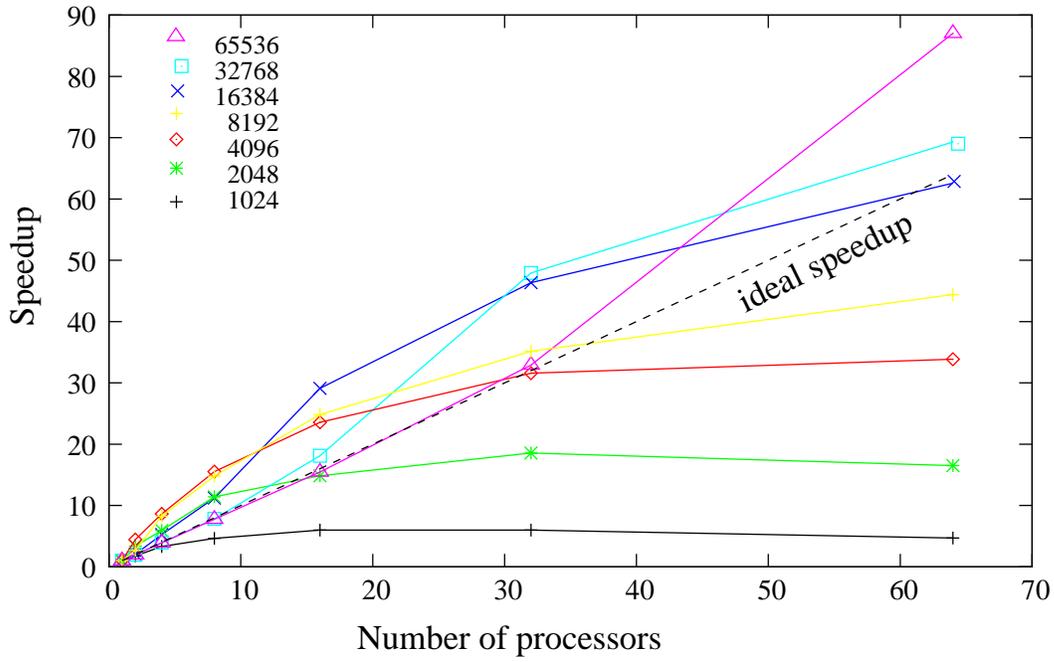,width=\columnwidth}
   \caption[]{
 Speedup for the LISA cluster as a function of the number of
 processors $p$ using 1 processor per node.  The various lines
 represent different particle numbers (see the legend in the top left
 corner). The diagonal dashed line give the ideal speedup. For $p
 \aplt 32$ the speedup of the code is super-ideal for $N \apgt 4096$.
\label{fig:speedup_LISA} }
\end{figure}

\begin{figure}
\psfig{figure=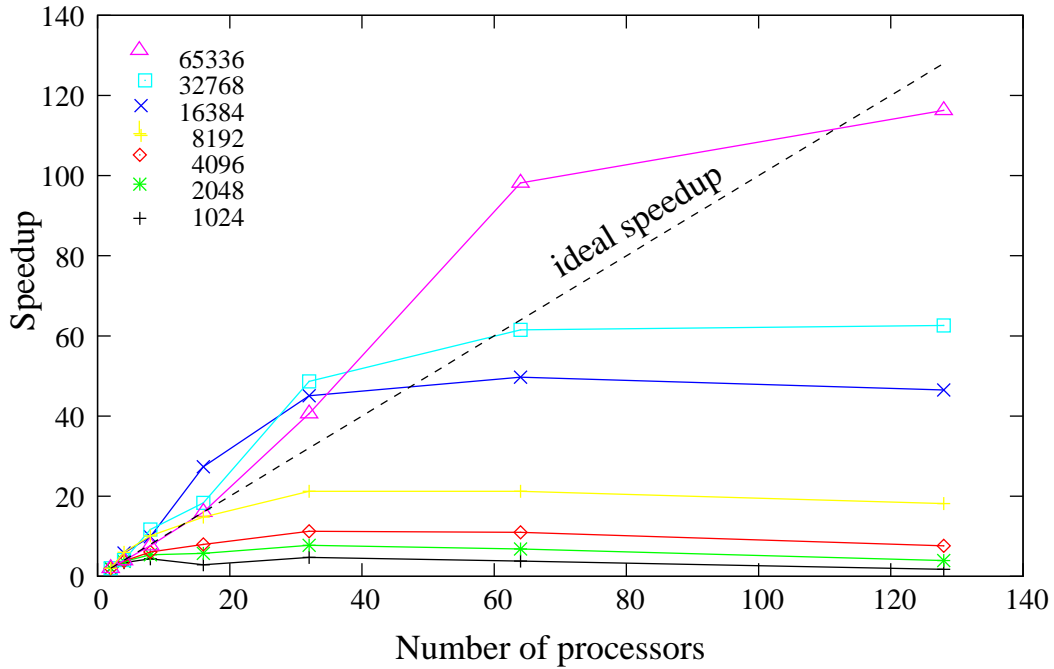,width=\columnwidth}
   \caption[]{
 Speedup for LISA using 2 processors per node.
\label{fig:speedup_LISA_2ppn} }
\end{figure}

\begin{figure}
\psfig{figure=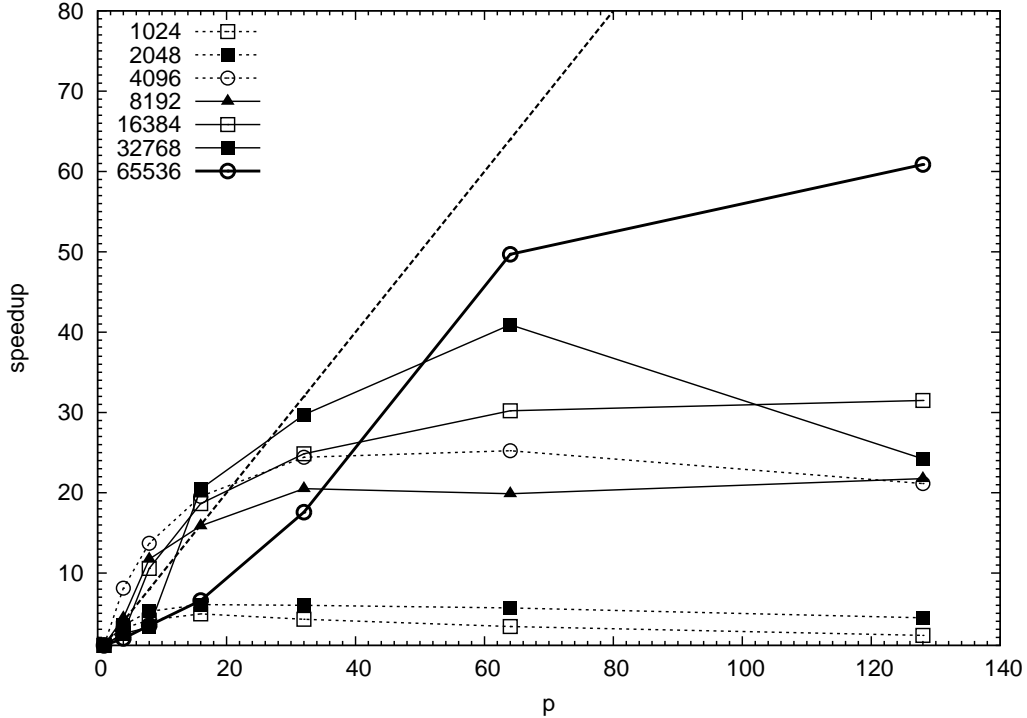,angle=270,width=\columnwidth}
   \caption[]{
 Speedup for the DAS-3 UvA cluster as a function of the number of
 processors $p$ using 4 processors per node (one process per core).
 The various lines represent different particle numbers (see the
 legend in the top left corner). The diagonal dashed line gives the
 ideal speedup.
\label{fig:speedup_DAS3} }
\end{figure}

\begin{figure}
\psfig{figure=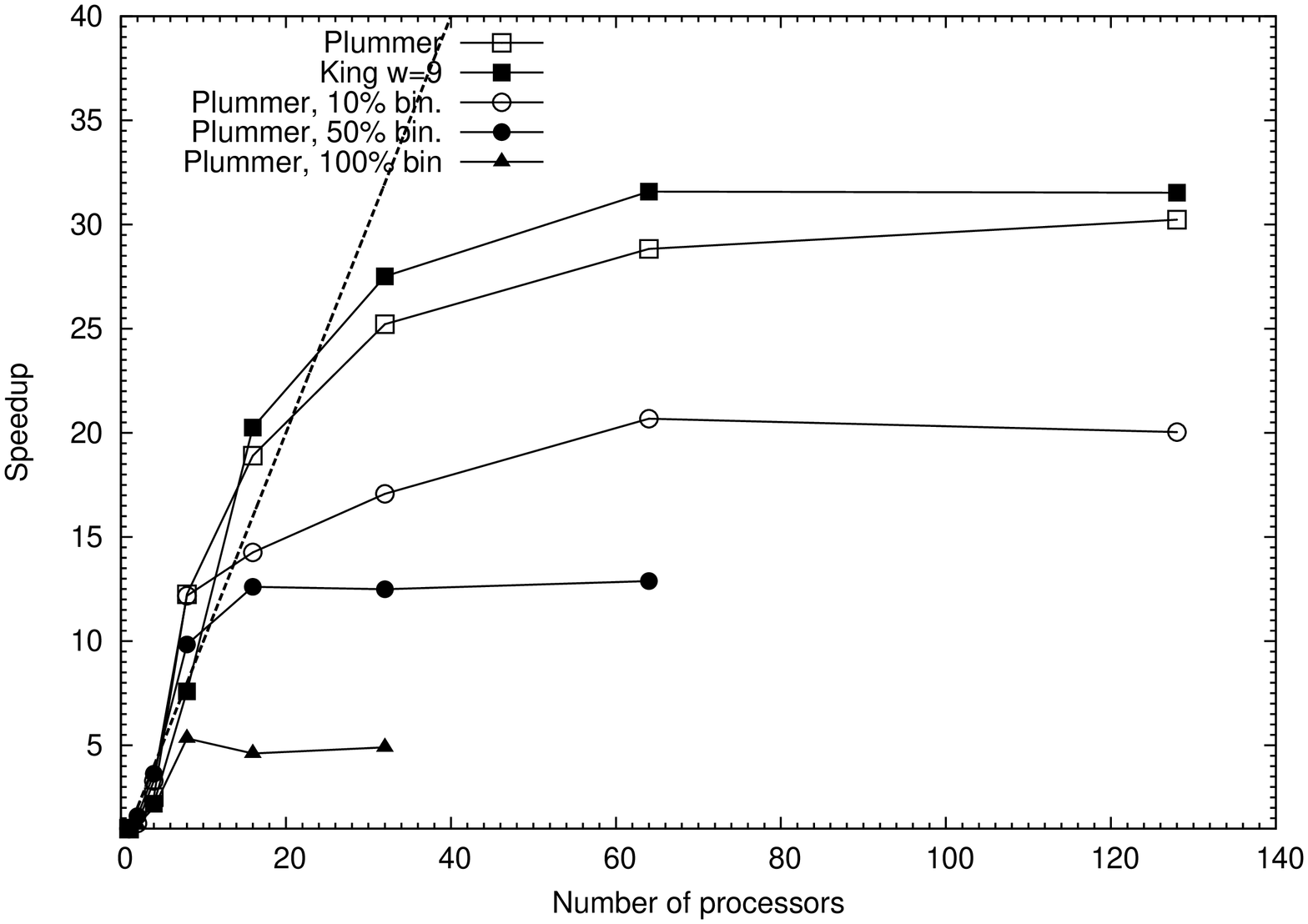,angle=270,width=\columnwidth}
   \caption[]{
 Speedup for $N=16384$ on the DAS-3 UvA cluster as a function of the
 number of processors $p$ using 4 processes per node (one process per
 core). The open squares indicate the Plummer distribution with single
 stars only (see also fig.\,\ref{fig:speedup_DAS3}).  The filled
 squares are simulated using a \cite{1966AJ.....71...64K} model with a
 dimension less depth of the central potential $W_0 = 9$, and with
 single stars only.  The other curves represent simulations with a
 Plummer sphere as initial conditions but with 10\% (circles), 50\%
 (bullets) and 100\% (triangles) primordial binaries of $10^4$kT.
\label{fig:speedup_DAS3_wbinaries} }
\end{figure}

For our performance measurements, we chose ASTER
(fig.\,\ref{fig:speedup_ASTER}), LISA (figs.\,\ref{fig:speedup_LISA}
and \ref{fig:speedup_LISA_2ppn}) and DAS-3 UvA
(figs.\,\ref{fig:speedup_DAS3} and \ref{fig:speedup_DAS3_wbinaries}).
In table\,\ref{tab:hardware} we present an overview of the computers
used for the performance measurements. The LISA supercomputer was used
in two different configurations, singe CPU and dual CPU.

\begin{table}
  \caption{Hardware specifications for the computers used for the 
performance tests.}
  \label{tab:hardware}
  \centering
  \begin{tabular}{lccc}
    \hline
                       & Aster & LISA & DAS-3\\
    \hline
    OS                 & Linux64 & Linux & Linux\\
    CPU                & Intel Itanium 2 & Intel Xeon & Dual Opteron DP 275\\
    CPU speed          & 1.3 GHz & 3.4 GHz & 2.2 GHz\\
    Memory/node [Gbyte]& 2 & 4 & 4\\
    Peak performance & 5.2 Gflop/s & 8.5 Tflop/s & 300 Glop/s\\
    $N_{\rm node}$     & 416 & 630 & 41 \\

    \hline
  \end{tabular}
\end{table}

\begin{figure}
\psfig{figure=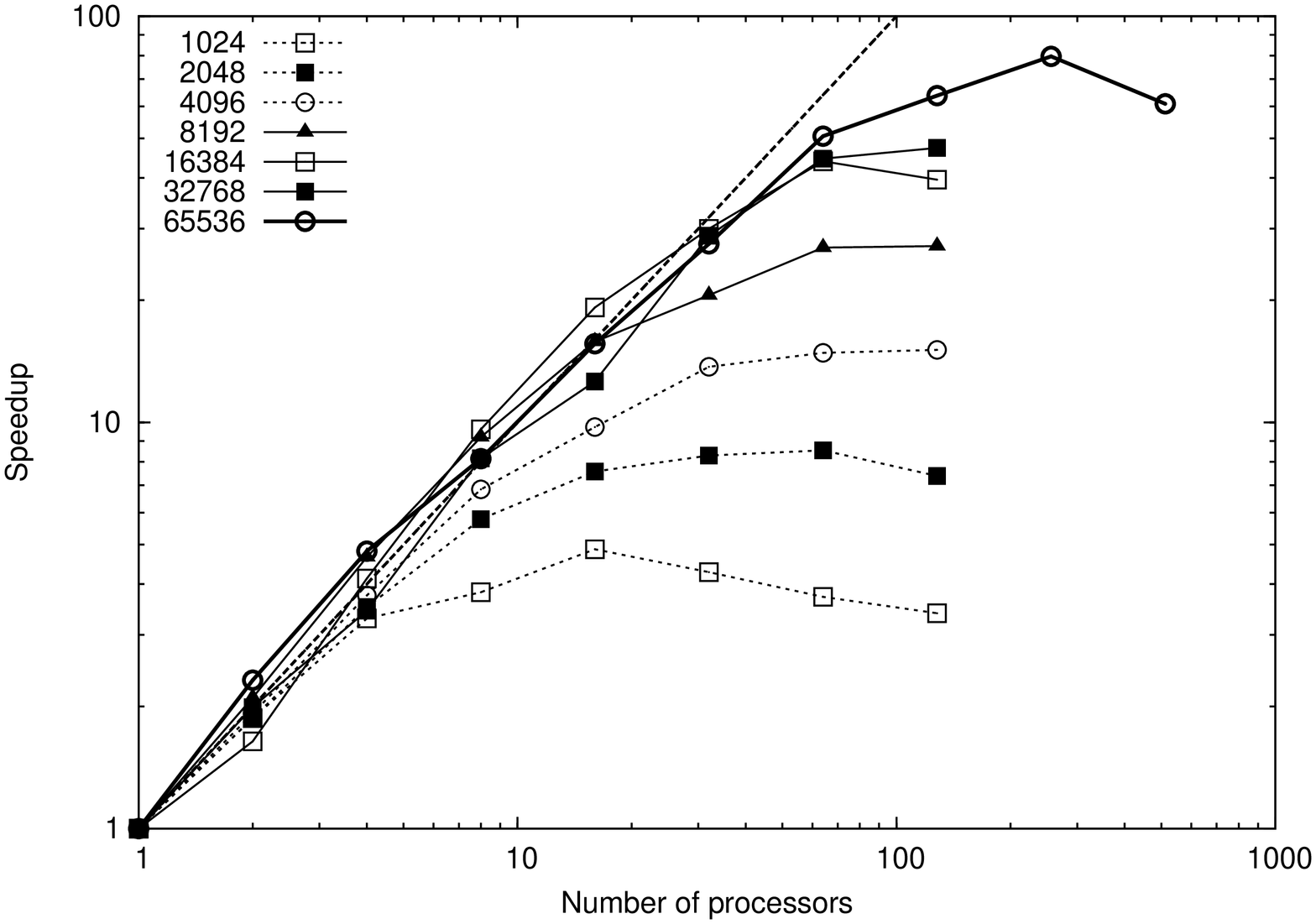,angle=270,width=\columnwidth}
   \caption[]{
 Speedup for the Huygens supercomputer.
\label{fig:speedup_Huygens} }
\end{figure}

The resulting speedup on the ASTER NUMA simulated shared memory
machine is presented in Fig.\,\ref{fig:speedup_ASTER}.  The results
with the new national supercomputer Huygens at SARA are presented in
Fig.\,\ref{fig:speedup_Huygens}.  The results for the LISA cluster
with one processor per node, and for the LISA with two processors per
node are presented in Fig.\,\ref{fig:speedup_LISA} and
\ref{fig:speedup_LISA_2ppn}, respectively.  In
Fig.\,\ref{fig:speedup_DAS3} we present the speed-up achieved on the
DAS-3 wide area computer.

We find a super-linear speedup on the LISA cluster using one Xeon
processor per node, but as expected the communications bottleneck
becomes quite noticeable for a larger number of processors. In
particular for simulations with relatively small $N$. This behavior is
due to a smaller memory footprint in individual processes, allowing the
nodes to benefit more from caching. 

In Fig.\,\ref{fig:speedup_Huygens} we present the speedup for {\tt
pkira} using the new national supercomputer Huygens. As with the other
architectures, the speedup is satisfactory up to about 100 processors
for $N \sim 64$k particles. 

Since {\tt pkira} is a production quality $N$-body code, we are able
to run simulations with a large fraction of primordial binaries and
with a large density contrast between the core and outskirts of
the cluster. The speedup of some test simulations with rather extreme
initial conditions are presented in
fig.\,\ref{fig:speedup_DAS3_wbinaries}.  For these tests we adopted
$N=16384$ in a Plummer sphere, as in Fig.\,\ref{fig:speedup_DAS3}.

The simulations with a King model with $W_0=9$ density profile (see
Fig.\,\ref{fig:speedup_DAS3_wbinaries}) show similar trends as the
simulations with a Plummer sphere, though with a slightly higher
speedup for 16--128 processors.  We conclude, based on these results
that the central concentration of the initial model has a slight
($\aplt 20$\%) positive effect on the speedup. This is mainly caused
by the slightly larger number of particles in each block time step.

Incorporating primordial binaries, however, has a dramatic effect on
the performance of the code.  In
Fig.\,\ref{fig:speedup_DAS3_wbinaries} we present the results of
simulations using a Plummer sphere but now with 10\%, 50\% and 100\%
primordial binaries of $10^4$ kT. Note that these simulations have a
proportionately larger number of stars than the models with single
stars.  For $\aplt 16$ processors we achieve a good speedup, but for a
larger number of processors the speeds up becomes a flat function of
the number of processors. The main bottlenecks for large binary
fraction runs on $p>16$ processors are the increased communication of
the perturber lists (see \S\,\ref{Sect:neighbors}) and the fact that
part of the perturber treatment is done sequentially.

\subsection{Comparison with a theoretical performance model}
\label{Sect:model}
To evaluate the performance of the parallel code for different
system sizes and processor numbers, we have implemented
a simple theoretical model and compared its predictions
with the timing data presented in Sec.\,\ref{Sect:performance}.

The total time needed to advance a block of particles of size $s$ can
be written as \cite{2007NewA...12..357H}
\begin{equation}
T_{\rm adv}\left(s\right) = T_{\rm host} + T_{\rm force} + T_{\rm comm},
\end{equation}
where $T_{\rm host}$ is the time spent on the host computer for the
predictor and corrector operations, $T_{\rm force}$ is the time spent
for the force computation, and $T_{\rm comm}$ is the time spent in
communication between processors.  

The time spent on the host for particle predictions and corrections
can be expressed as $T_{\rm host} = T_{\rm pred} + T_{\rm corr} =
\tpr\,N + \tcr\,s$, where $\tpr$ and $\tcr$ represent the time
required to perform, respectively, the prediction and the correction
operation on one particle. We note that the predictor is performed for
all particles while the corrector is performed for the particles in
the block being advanced.

The time needed for one computation of the force between two particles
is indicated by $\tau_f$ and consequently, the time to compute the
force on a block of particles is given by $T_{\rm force} =
\tau_f\,N\,s/p$; each processor contributes to the computation of the
force exerted on the particles in the block by the its local $N/p$
particles.

The time for communication $T_{\rm comm}$ is given by the sum of times
spent in each MPI call, and is dominated by two calls to the function
{\tt MPI\_Allreduce} and three calls to the function {\tt
MPI\_Allgatherv}.  We adopt the following models for the MPI
functions:
\begin{eqnarray}
T_{\tt MPI\_Allgatherv} &=& \left( \alpha + \beta\,x\right) \log_2\,p ,
\nonumber \\
T_{\tt MPI\_Allreduce} &=& \left( \delta + \gamma\,x\right) \log_2\,p
\end{eqnarray}
where $x$ represents the size of transferred data measured in bytes
and $\alpha = 1.2\times10^{-5} \,\rm s$, $\beta = 1.3\times10^{-9}\, \rm
s/byte$, $\delta = 1.0\times10^{-5}\, \rm s$, $\gamma =
5.0\times10^{-9}\, \rm s$ are the parameters obtained by fitting
timing measurements on the LISA cluster.

This analysis shows how the time for advancing a block of particles
depends on the speed of calculation of each processor, on the latency,
and on the bandwidth of communication among the processors.  The
parameters that appear in the performance model were measured on the
LISA cluster and resulted in: $\tpr = 5.6\times 10^{-8}\,\rm s$, $\tcr
= 1.0\times 10^{-6}\,\rm s$, $\tau_f = 1.2\times 10^{-7}\, \rm s$.
The overall performance of our implementation is not optimal, and in
the absence of special hardware much better performance could be
obtained by adopting the low-level implementation of the force
operation as was presented recently by \cite{2006NewA...12..169N}.

\begin{figure}
\psfig{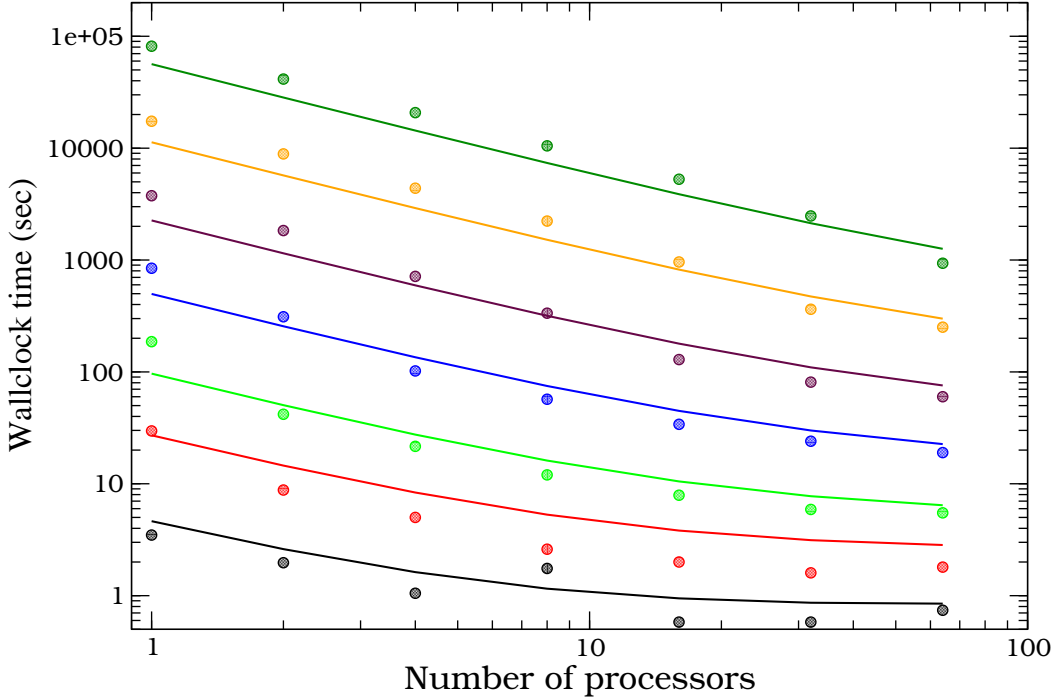}
   \caption{Comparison between the predictions by the theoretical
            model and the timing measurements on the LISA cluster for
            the integration of Plummer models over a fraction of one
            eighth of an $N$-body unit. The solid lines represent the
            predictions by the model while the data points indicate
            the timing measurements.  Different lines refer to Plummer
            models with different numbers of particles, increasing
            from bottom ($N=1024$) to top ($N = 65536$).}
\label{fig:model} 
\end{figure}

To predict the total time for the integration of a system over one
$N$-body unit (or any other physical time), it is necessary to know
the block size distribution for the model under consideration.  In the
case of the Plummer model, the total execution time over one $N$-body
time unit can be estimated by considering the average value of the
block size $\bar{s}$ and the total number of integration
steps $n_{\rm steps}$ in one $N$-body unit,
\begin{equation}
\label{eq:tsnb}
T_N = T_{\rm adv}(\bar{s})\,n_{\rm steps}\,.
\end{equation}
We have measured the average block size and the number of block time
steps in one $N$-body unit for Plummer models of different $N$ and
applied the theoretical model to the prediction of the total execution
times for the same models. Fig.\,\ref{fig:model} shows a comparison
between the predicted times (solid lines) and timing measurements
(data points) conducted on the LISA cluster for Plummer models of
different numbers of particles.

The model and simulation agree best for 4096 - 16384 particles. It
tends to overestimate the wallclock time for small, and underestimates
for larger numbers of particles.  The deviations of the predicted
times for $N \aplt 4096$ are mainly caused by the small block sizes,
which renders the average $\bar{s}$ a poor indicator of the global
behavior of the system.  For large $N$, the model tends to
underestimate the total time, because the model only takes the times
required by the predictor, corrector, force calculation and MPI
communication into account, ignoring the other operations, like
sorting and broadcasting the neighbor list, and treatment of binaries.
Overall, we consider the comparison over the entire range of $N$
reasonable, given the relative simplicity of the performance model.

\subsection{Pkira performance on the grid}
\label{Sect:grid}
We tested {\tt pkira} in parallel across clusters, using the same
initial conditions as for the local performance test with 65536
particles. We have executed the simulation, using up to 128 processes,
in parallel across 1,2 and 4 clusters respectively. Across clusters,
we made use of regular networking.

\begin{table}
  \caption{Hardware specifications for the DAS-3 computing sites used
for the grid performance tests. All nodes have 4 GB of memory and the
latency between the sites is up to 2.2 ms.}
  \label{tab:hardware}
  \centering
  \begin{tabular}{lcccc}
    \hline
                       & UvA & VU & Leiden U. & MultimediaN\\
    \hline
    OS                 & Linux & Linux & Linux & Linux\\
    CPU                & Dual Opt.\, DP 275 & Dual Opt.\, DP 280 & Opt.\, DP 252 & Opt.\, DP 250\\
    CPU speed          & 2.2 GHz & 2.4 GHz & 2.6 GHz & 2.4 GHz\\
    $N_{\rm node}$     & 41 & 85 & 32 & 46\\

    \hline
  \end{tabular}
\end{table}

We find that the runs across clusters are slower than single-cluster
runs with the same number of processors, due to the long network
latency (of up to 2ms) and relatively low bandwidth. The efficiency, however, is
above 0.9 when we compare a run with 32 processes on one cluster, or
over two clusters with 16 processes each. Running {\tt pkira}
over two clusters with 32 processes each, rather than 32 processes on
a single cluster, provides a speedup of $\sim1.9$.

\begin{figure}
\psfig{figure=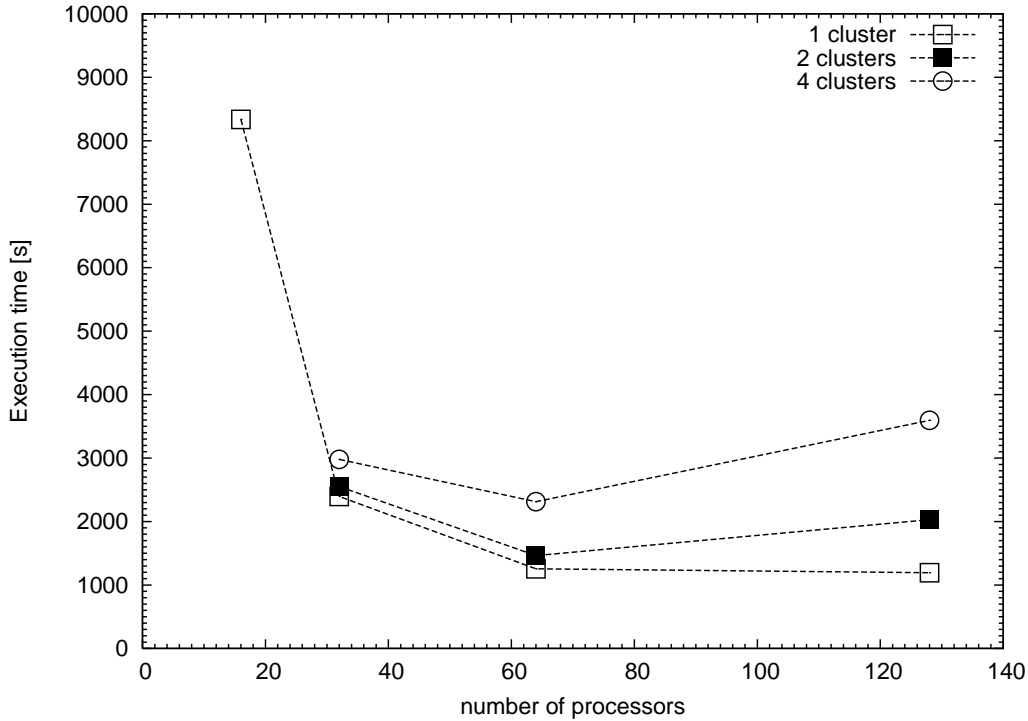,angle=270,width=\columnwidth}
   \caption[]{
 Performance for the DAS-3 wide area computer using 1 to 4 sites
 (clusters) as a function of the total number of processes $p$ using
 one process per CPU core.  The various lines represent the number of
 clusters used (see the legend in the top right corner).
\label{fig:grid_DAS3} }
\end{figure}

The results of our simulations on the Grid are summarized in
Fig.\,\ref{fig:grid_DAS3}, where we show the execution time for a
Plummer sphere with $N=16384$ as a function of the number of
processors in the testbed. The same number of processors are divided
across a single cluster (open squares), two clusters (filled squares)
and 4 clusters (circles). These simulations were performed on the
DAS-3.  For $\aplt 60$ processors the performance across two sites is
not much affected by the long baseline between the two clusters. For 4
clusters the execution time is dominated by the farthest site, which
has the longest latency and somewhat lower bandwidth. For grid-based
simulations it appears useless to go beyond 60 processors for $N \leq
16384$, since the total performance drops, resulting in a lower total
execution time.  Note that for $N=16384$ on 128 processors at a single
site doen not result in a speedup either. If relatively few processors
are available on a single site, it may be worth considering using a
grid, if that allows the user to utilize a larger number of
processors, as long as the number of processors $\aplt 60$.

%----------------------------------------------------------------------
\section{Discussion and conclusions}

We report the results of our parallelization on the {\tt kira}
$N$-body integrator in the {\tt starlab} package. In order to maintain
a platform independent source code we opted for the Message Passing
Interface standard.  The parallelization scheme is standard
$j$-parallelization with a copy of the entire system on each node to
guarantees that the perturber lists are consistent with the serial
version of the code.  In this way we achieve that the numerical
results are independent of the number of processor units.  We had to
include an additional broadcast step to guarantee that the neighbour
list is consistent between the sequential and the parallel code. This
extra overhead, is particularly noticeable for small numbers of
particles, and does not seriously hamper the scaling of the code with
the number of processors. For simulations with a large number of
binaries or when running on clusters with more than 128 processors we
also found a drop in performance.

The code scales well in the domain tested, which ranges from 1024 to
65536 stars on 1 to 128 processors.  A super linear speed-up was
achieved in several cases for 4096--32768 stars on 16--64 processors.

The overall performance for clusters with single stars is rather
satisfactory, irrespective of the density profile of the star
cluster. But for primordial binaries the performance drops
considerably. With 100\% binaries the performance of a large
simulation on many processors hardly results in a reduction of the
wall-clock time compared to a few processors. Runs with a large number
of primordial binaries still benefit from parallelization for $\aplt
16$ processors.  This performance hit in the presence of binaries is
mainly the results of the additional communication required to
synchronize the perturber list and of the sequential part in the
binary treatment.

\section*{Acknowledgments}

We are grateful to Douglas Heggie, Piet Hut and Jun Makino for
discussions.  This research was supported in part by the Netherlands
Supercomputer Facilities (NCF), the Netherlands Organization for
Scientific Research (NWO grant No. 635.000.001 and 643.200.503), the
Netherlands Advanced School for Astronomy (NOVA), the Royal
Netherlands for Arts and Sciences (KNAW), the Leids Kerkhoven-Bosscha
fonds (LKBF) by NASA ATP grant NNG04GL50G and by the National Science
foundation under Grant No. PHY99-07949 and DEISA.  The calculations
for this work were done on the LISA workstation cluster, the ASTER
supercomputer, de DAS-3 wide area computer in the Netherlands and the
MoDeStA computer in Amsterdam, which are hosted by SARA computing and
networking services, Amsterdam.  AG is supported by grant NNX07AH15G
from NASA.

%\input /home/spz/latex/lib/bib/references
%\bibliographystyle{mn2e}
%\bibliography{biblio}

\end{document}